\documentclass[conference]{IEEEtran}
\IEEEoverridecommandlockouts
\usepackage{cite}
\usepackage{amsmath,amssymb,amsfonts}
\usepackage{algorithmic}
\usepackage{algorithm}  
\usepackage{graphicx}
\usepackage{textcomp}
\usepackage{xcolor}
\usepackage{float} 
\usepackage{color}
\usepackage{multirow}
\usepackage{array}
\usepackage{booktabs}
\usepackage{multirow}
\usepackage{fancyhdr} 
\def\BibTeX{{\rm B\kern-.05em{\sc i\kern-.025em b}\kern-.08em
    T\kern-.1667em\lower.7ex\hbox{E}\kern-.125emX}}

\usepackage{fancyhdr}
\renewcommand{\headrulewidth}{0pt}
\renewcommand{\footrulewidth}{0pt}

\begin{document}

\title{ Decentralized Microgrid Energy Management: A Multi-agent Correlated 
Q-learning Approach\\
}
\author{\IEEEauthorblockN{Hao Zhou, and Melike Erol-Kantarci, Senior Member, IEEE}
\IEEEauthorblockA{\textit{School of Electrical Engineering and Computer Science} \\
\textit{University of Ottawa}\\
Emails:\{hzhou098, melike.erolkantarci\}@uottawa.ca}}

\maketitle

\thispagestyle{fancy} %
      \lhead{} 
      \chead{Accepted by 2020 IEEE International Conference on SmartGridComm, 978-1-7281-6127-3/20/\$31.00~ \copyright2020 IEEE 
 } 
      \rhead{} 
      \lfoot{} 
      \cfoot{\thepage} 
      \rfoot{} 
      \renewcommand{\headrulewidth}{0pt} 
      \renewcommand{\footrulewidth}{0pt} 
\pagestyle{fancy}
\cfoot{\thepage}

\begin{abstract}
Microgrids (MG) are anticipated to be important players in the future smart grid. For proper operation of MGs an Energy Management System (EMS) is essential. The EMS of an MG could be rather complicated when renewable energy resources (RER), energy storage system (ESS) and demand side management (DSM) need to be orchestrated. Furthermore, these systems may belong to different entities and competition may exist between them. Nash equilibrium is most commonly used for coordination of such entities however the convergence and existence of Nash equilibrium can not always be guaranteed. To this end, we use the correlated equilibrium to coordinate agents, whose convergence can be guaranteed. In this paper, we build an energy trading model based on mid-market rate, and propose a correlated Q-learning (CEQ) algorithm to maximize the revenue of each agent. Our results show that CEQ is able to balance the revenue of agents without harming total benefit. In addition, compared with Q-learning without correlation, CEQ could save 19.3\% cost for the DSM agent and 44.2\% more benefits for the ESS agent.  
\end{abstract}


\section{Introduction}

Microgrids (MG) are becoming promising solutions to enhance the efficiency, resilience and flexibility of future smart grid \cite{b1}. Due to the integration of renewable energy resources and energy storage system (ESS), MGs are able to share their stored energy with each other to enhance reliability \cite{b2}.  Resilient MGs are even expected to perform well under a catastrophic event to serve critical services as envisioned in \cite{b3}.  The EMS of MG, which is the key of MG operation, could be centralized or decentralized. Centralized EMS needs to deal with the global information of MG, which will increase the complexity \cite{b4}. Therefore decentralized EMS is regarded as the future trend despite its control related challenges. 

Meanwhile, the recent years have witnessed the great potential of AI techniques, which provides a new opportunity to improve the MG operation \cite{b5,b6,b7}. As a model-free algorithm, Q-learning can avoid the complexity of building a detailed optimization model. Q-learning is used in \cite{b5} to build a decentralized EMS in a MG where the benefits of agents is balanced by Nash equilibrium. \cite{b6} extended the reinforcement learning to MGs, which aimed to minimize the power loss using Bayesian reinforcement learning. In \cite{b7}, the authors focus on the energy trading between MGs, where reinforcement learning is combined with Stackelberg game to find the Nash equilibrium.

In practice, the agents that control generation, storage, demand in a MG may belong to different entities, and they may be competing with each other for maximizing their revenues. In such a case, the coordination of agents are crucial for proper MG operation. In the literature, Nash equilibrium is widely used to coordinate the operation of agents \cite{b5,b7,b8,b9}. In \cite{b5}, the interaction between agents became optimal when Nash equilibrium is reached, which means no agent can improve its expected reward by changing current strategy. In \cite{b9}, Nash equilibrium is found in a distributed way, where no central controller is needed. However, Nash equilibrium is not guaranteed to converge or reach a global-optimal result, and some games have multiple Nash equilibriums. \cite{b9,b10}.  In some cases, the Nash equilibrium is not unique or do not even exist \cite{b11}. 
 
To this end, we use the correlated equilibrium to coordinate agents, which is more general than Nash equilibrium. Firstly, the correlated equilibrium allows for dependencies among agents’ strategies, while the actions in Nash equilibrium should be independent. Secondly, compared with iterative method which is generally used for finding Nash equilibrium, correlated equilibrium could be easily found by linear programming. The convergence and existence of correlated equilibrium is further proved in \cite{b10}. 

The main contribution of this paper is that we propose a multi-agent based correlated Q-learning (CEQ) algorithm for MG energy management which is implemented in a decentralized manner. The simulation results show that CEQ is capable of coordinating agents. The result shows the DSM agent could save as much as 19.3\% of cost, and the ESS agent
could earn 44.2\% more benefit. Moreover, we compare the result with centralized EMS, where all agents belong to one aggregator, and we get the same total revenue which implies CEQ maintains the benefits of a centralized algorithm.

The rest of this paper is organized as follows. Section II presents the related work. Section III introduces our community MG system model, and Section IV introduces the proposed CEQ algorithm. Section V shows simulation results and finally, Section VI concludes the paper. 

\section{Related Work}

Energy management is the key for efficient operation of MGs. A controller coordinates the operation of different agents and optimizes the overall efficiency. With the integration of RER, ESS, DSM and other agents, the complexity of MG energy management increases greatly. Based on AI techniques, learning-based methods become promising solutions for MG energy management. These have been investigated in a few recent studies which are summarized below \cite{b12,b13,b14,b15,b16}.

In \cite{b12}, fuzzy Q-learning is used for an isolated MG with multiple agents where the agents share their state variables for coordination. \cite{b13} uses a decision tree to chose actions for MG agents where the large training episodes are divided into small pieces, and Q-table is transferred between small episodes to speed convergence. Reinforcement learning and deep neural network is combined in \cite{b14} to conduct the energy management of multiple MGs. Furthermore, CEQ is used for dynamic transmission control of sensor networks in \cite{b15}, where sensors learn the correlated equilibrium policy independently. In \cite{b16}, CEQ is applied for smart generation control of power grids, where each area has one independent generation agent. 

Considering the potential advantage of CEQ, this paper extends the application of CEQ for smart grid and MG integration. Different than [16], this paper mainly investigates the CEQ for MG energy management. In [16], 
 the reward function is the same for each agent and the objective is solely generation control. However, the interactions between agents are more complicated in MGs since generation, storage and demand is involved. Therefore, one needs to define the reward function for each agent based on the their own characteristics. To the best of our knowledge, this is the first time CEQ is used for MGs. 

\section{Community Microgrid System Model}

 In this section, we will introduce the demand side management model, the PV model and the energy storage system model. Based on mid-market rate, an improved energy trading model is built.
\subsection{Demand Side Management Model}
The user demand is divided as crucial load and deferrable load. The demand side management (DSM) agent does not interfere with the crucial load, e.g., lighting and cooking devices. However deferrable loads, such as a dish washer or a water heater, can be deferred to another time by the DSM agent. 

For $n$ sets of deferrable devices, the state is described as: 
\begin{equation} \label{eu_eqn}
\vec a_{t}=[a_{t,1},a_{t,2},\dots,a_{t,i},\dots, a_{t,n}]
\end{equation}
where $a_{t,i}=1$ represents turning on the devices in set $i$ at time $t$; and $a_{t,i}=0$ represents turning off. 

The power demand of DSM agent at time $t$ is:
\begin{equation} \label{eu_eqn}
P^{DSM}_{t} =\sum_{i=1}^{n}P_{i}a_{t,i}
\end{equation}
where $P_{i}$ is the power of deferrable devices set $i$. We assume an average power consumption for the duration  $[t,t+1]$.

The deferrable devices must be serviced before a certain time limit, which means devices can not be deferred longer than this limit. The waiting time of devices is described as:
\begin{equation} \label{eu_eqn}
\vec w_{t}=[w_{t,1},w_{t,2},\dots,w_{t,i},\dots, w_{t,n}]
\end{equation}
where $w_{t,i}=1$ represents the devices set $i$ that is still under service at time $t$; $w_{t,i}=0$ represents this devices set has been serviced or its turn has not come. We assume that if one device reaches its time limit and it has not been serviced, then it will be turned on mandatorily. 

\subsection{PV Model}
In this paper, we assume the PV power can be  predicted with an acceptable error \cite{b17}.
\begin{equation} \label{eu_eqn}
P^{PV}_{t} =\hat{P}^{PV}_{t}
\end{equation}
where $\hat{P}^{PV}_{t}$ is the predicted PV power.

\subsection{Energy Storage System Model}
In this research, we consider centralized ESS in the community MG. We use $q_{t}$ to denote the state of ESS at time $t$. There are two discharging levels, denoted as $q_{t}$ equals 0.5 and 1, to enhance flexibility. This means the ESS can be either fully discharged or half discharged. The power of ESS is: 
\begin{equation} \label{eu_eqn}
P^{ESS}_{t} =P^{char}q_{t}
\end{equation}

\begin{equation} \label{eu_eqn}
 q_{t}=\left\{
\begin{array}{rcl}
-1     &      &  charge\\
 0     &      &  unchanged\\
0.5    &      &  discharge\\
 1     &      &  discharge
\end{array} \right.
\end{equation}
where  $P^{char}$ is a constant charging power. 

The state of charge (SOC) of ESS is updated according to: 
\begin{equation} \label{eu_eqn}
SOC_{t+1} =SOC_{t}-\frac{P^{char}}{C^{ESS}}q_{t}
\end{equation}
where $C^{ESS}$ is the capacity of ESS.

\subsection{Energy Trading Model}
Our energy trading model is presented in Fig.1. According to our model, DSM agent could choose ESS or main grid as its energy source. ESS agent could use PV or grid power to charge, and it can sell its energy to the DSM agent or the main grid. We make the following assumptions to build the energy trading model:

Assumption 1: Compared with the main grid, PV power has a lower price: $p^{PV}<p^{bgrid}_{t}$. Considering the rationality of main grid, $p^{sgrid}_t<p^{bgrid}_t$.

Assumption 2: Surplus PV power is only available when PV power is more than the crucial load. Note that energy trading of crucial load is out of the scope of this paper.

Assumption 3: We assume ESS is unable to charge and discharge at the same time. 

Assumption 4: If DSM is unable to consume all the ESS energy, ESS could sell the rest energy to main grid. 

The mid-market rate has been generally used in P2P energy trading research as in \cite{b18}. We use $p^{mid}_t$ to represent the price of ESS selling electricity to DSM.
\begin{equation} \label{eu_eqn}
p^{mid}_t=\frac{p^{sgrid}_t+p^{bgrid}_t}{2}
\end{equation}
 where $p^{sgrid}_t$ and $p^{bgrid}_t$ represent the price of selling power to the main grid and buying power from the main grid separately.  

However, in a competitive situation, it is reasonable to assume one agent may take risks for a higher profit, which is denoted as a threat strategy. The strategy combination and results are presented in Table 1. If one agent chooses a threat strategy and other agent still cooperates, the former agent will earn more benefit. If both of them choose threat strategy, then cooperation breaks up, both agents will exchange power with main grid.
In this paper, we will use correlated equilibrium to coordinate the actions of agents, avoid conflict and balance the revenue. 
\begin{figure}[t]
\centering
\includegraphics[width=7cm,height=3.4cm]{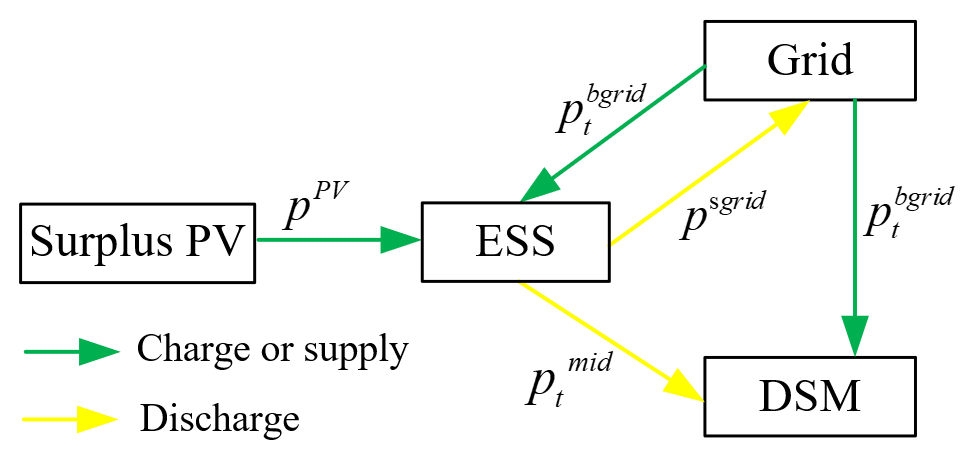}
\caption{Energy trading model.}
\label{fig}
\end{figure}

\begin{table}[t]
\begin{center} 
\caption{Strategy combination of [players}
\renewcommand\arraystretch{2}
\centering
\begin{tabular}{|p{0.6cm}<{\centering}|p{0.9cm}<{\centering}|p{2.9cm}<{\centering}|p{2.9cm}<{\centering}|}
\hline
\multicolumn{2}{|c|}{Value of }&\multicolumn{2}{|c|}{ESS agent}\\ 
\specialrule{0em}{0pt}{0pt}
\cline{3-4}
\multicolumn{2}{|c|}{middle price} & Cooperation& Threat \\ 
\hline
\multirow{2}*{\shortstack{DSM\\Agent}}& \multirow{1}*{\shortstack{Cooper-\\ation}} &$0.5p^{sgrid}+0.5p^{bgrid}$ & $0.25p^{sgrid}+0.75p^{bgrid}$ \\
\cline{2-4}
& Threat & $0.75p^{sgrid}+0.25p^{bgrid}$& \multirow{1}*{\shortstack{Cooperation\\breaks}}\\
\hline
\end{tabular}
\end{center} 
\end{table}

\subsection{Problem Formulation}

Both ESS and DSM agent make decisions independently to optimize their own operation. The optimization objective of the ESS agent is maximizing its revenue:
\begin{equation} \label{eu_eqn}
max(P^{char}\sum_{t=1}^{T}q_{t}p^{clear}_{t})
\end{equation}
where $T$ is simulation period; $p^{clear}_{t}$ is the price of ESS agent buying/ selling electricity, which is known as the market clearing price \cite{b13}. Clearing price denotes an ideal situation, where the demand equals the supply and no shortage or surplus exist in the market. $P^{char}$ and $q_{t}$ have been defined in equation (5).

On the other hand, the optimization objective of DSM agent is minimizing cost:
\begin{equation} \label{eu_eqn}
min(\sum_{t=1}^{T}P^{DSM}_{t}p^{buy}_{t})
\end{equation}
where $p^{buy}_{t}$ is the price of buying power at time $t$.

The optimization has to obey following constraints:
\begin{equation} \label{eu_eqn}
P^{PV}_{t}+P^{grid}_{t}+P^{ESS}_{t}=P^{DSM}_{t}
\end{equation}
\begin{equation} \label{eu_eqn}
SOC_{min}\leq SOC_{t} \leq SOC_{max}
\end{equation}
\begin{equation}
a_{t,i}\leq w_{t,i}
\end{equation}

Equation (11) denotes the energy balance constraint, where $P^{grid}_t$ is the power from the main grid. Equation (12) is the SOC constraint, where $SOC_{min}$ and $ SOC_{max}$ are lower and upper bound of ESS. Equation (13) is the DSM constraint, which means only devices that have not been serviced could be turned on.  

\section{Correlated Q-learning Algorithm}
In this paper, we propose a correlated Q-learning (CEQ) based algorithm to coordinate the operation of agents.
CEQ is a multi-agent reinforcement learning algorithm. The coordination between agents is achieved by exchanging the state-action value matrix, which means it could be implemented in a decentralized way. We will introduce the system state, actions, rewards  and correlated equilibrium in this section.

\subsection{State and Actions}
The system state is defined as: 
\begin{equation} \label{eu_eqn}
s=\left\{ t, SOC,\vec w  \right\}  
\end{equation}

The actions of DSM  and  ESS agent are:
\begin{equation} \label{eu_eqn}
a^{DSM}=\left\{ y,\vec a  \right\}  
\end{equation}
\begin{equation} \label{eu_eqn}
a^{ESS}=\left\{ x, u  \right\}  
\end{equation}
where $y$ belongs to a set of two choices: buying electricity from ESS or main grid; $x$ belongs to a set of two choices: selling power to main grid or DSM, $u$ is the set of $q_{t}$ in eq. (5).

The Q value spaces of DSM and ESS agent are:
\begin{equation} \label{eu_eqn}
Q_{DSM}=\left\{ a^{ESS},a^{DSM},  t,SOC,\vec w \right\} 
\end{equation}
\begin{equation} \label{eu_eqn}
Q_{ESS}=\left\{ a^{ESS},a^{DSM},  t,SOC,\vec w \right\}  
\end{equation}

\subsection{Reward}
The cost of ESS agent includes buying power from PV or main grid,  and the benefit comprises of selling power to grid or DSM. Meanwhile, DSM agent aims to minimize the cost by adjusting operation time of deferrable devices and choose a lower price energy source. The reward can be formulated for the following two cases that correspond to ESS charging or discharging.

1)  At time $t$, if ESS agent chooses to charge, the instant reward is: 
\begin{equation} \label{eu_eqn}
r^{ESS}_{t}=P^{ESS}_{t}(\alpha p^{PV}_{t}+(1-\alpha)p^{bgrid}_{t})
\end{equation}
where $\alpha$ denotes the proportion of buying power from PV $(0\le\alpha\le 1)$.

ESS agent will prefer PV power because of a lower price, but surplus PV power may not be enough for ESS charge. As a liner program problem, we give the following equation of  $\alpha$:
\begin{equation} \label{eu_eqn}
\alpha=\frac{P^{PV}_{t}-P^{crl}_t}{\left| P^{ESS}_{t} \right|}
\end{equation}
where $P^{crl}_t$ is the crucial load.

Meanwhile, the DSM can only get power from the main grid. The reward of DSM agent at time $t$ is:
\begin{equation} \label{eu_eqn}
r^{DSM}_{t}=-P^{DSM}_{t}p^{bgrid}_{t}
\end{equation}

2) If ESS agent chooses to discharge, the reward of ESS agent is:
\begin{equation} \label{eu_eqn}
r^{ESS}_{t}=P^{ESS}_{t}(\beta p^{mid}_{t}+(1-\beta)p^{sgrid}_{t})
\end{equation}
where $\beta$ denotes the proportion of selling power to DSM $(0\le\beta\le 1)$. According to equation (8), the ESS will try to sell power to DSM agent for a higher price. Note that DSM may be  unable to buy all ESS power, therefore we give the following equation:
\begin{equation} \label{eu_eqn}
\beta=\frac{P^{DSM}_{t}}{ P^{ESS}_{t}}
\end{equation}
The reward of DSM agent is:
\begin{equation} \label{eu_eqn}
r^{DSM}_{t}=-P^{DSM}_{t}(\gamma p^{mid}_{t}+(1-\gamma)p^{bgrid}_{t})
\end{equation}

Similarly, denoting the proportion of buying power from ESS by:
\begin{equation} \label{eu_eqn}
\gamma=min(\frac{P^{ESS}_{t}}{ P^{DSM}_{t}},1)
\end{equation}

\subsection{Correlated Equilibrium}
The aim of ESS and DSM agent are both to maximize their reward in a simulation epoch. 
For the agent $i$, with the initial state $s_{0}$, the expected accumulated discounted reward under a policy $\pi$ is:
\begin{equation} \label{eu_eqn}
V^{\pi}_{i}(s) =E_{\pi}[\sum_{n=0}^{\infty}\theta^{n} r_{i}(s_{n},a_{n})|s=s_{0}]
\end{equation}
where  $\theta$ is the reward discount factor.

For a specific state $s$ and action $a$, we define the state-action value:
\begin{equation} \label{eu_eqn}
Q^{\pi}_{i}(s,a) = r_{i}(s,a)+\theta\sum_{s'\in S}P(s'|s,a)V^{\pi}_{i}(s)
\end{equation}
Then  we have the following relationship:
\begin{equation} \label{eu_eqn}
V^{\pi}_{i}(s) = \sum_{a\in A}\pi_{x}(a)Q_{i}(s,a)
\end{equation}
 
 In the CEQ algorithm, agents chose a joint optimal action by exchanging Q value matrix, and we assume both agents know the possible actions of each other. The joint action of agents are chosen according to correlated equilibrium:
\begin{equation} \label{eu_eqn}
\begin{split}
\sum_{a_{-i}\in A_{-i}}\pi_{x}(a_{-i},a_{i})Q_{i}(s,a_{-i},a_{i})\geq\\ 
\sum_{a_{-i}\in A_{-i}}\pi_{x}(a_{-i},a_{i})Q_{i}(s,a_{-i},a'_{i})
\end{split}
\end{equation}
where the $a_{i}$ and $a'_{i}$ denote the action of agent $i$ in correlated equilibrium and non-equilibrium, $a_{-i}$ denotes the actions of agents except agent $ i $,  $A_{-i}$ denotes the action set of agents except $ i $. Meanwhile, equation (29) denotes that the joint-optimal action is chosen by exchanging Q-values, which means the private information of each agent could be well protected. 

The policy is improved according to $\epsilon$-greedy policy:

\begin{equation} \label{eu_eqn}
\pi_{i}(s) =\left\{
\begin{array}{rcl}
  random &   rand\leq\epsilon\\
equation(29)  &  rand>\epsilon\\
\end{array} \right.
\end{equation}
Where $\epsilon$ is a small value between 0 and 1; $rand$ is a random number between 0 and 1. 

It is worth noting that ESS will only choose eligible actions, e.g., the charge action is eliminated if $SOC$=1. The CEQ algorithm is summarized in Algorithm 1.

\begin{algorithm}[!h]
	\caption{Correlated Q-learning}
	\begin{algorithmic}[1]
		\STATE \textbf{Initialize:} microgrid and Q-learning parameters
		\FOR{$j=1$ to $episode$}
		\STATE Reset state s 
		\FOR{$t=1$ to $T$}
		\STATE Calculate eligible $a_{t}^{DSM}$, $a_{t}^{ESS}$ according to state $s$
		\IF{$rand<\epsilon$}
		\STATE Randomly choose $a_{t}^{DSM}$, $a_{t}^{ESS}$  
		\ELSE
		\STATE Agents exchange current state-action matrix \STATE Find equilibrium by equation (29) and get optimal joint action $a_{t}^{DSM}$, $a_{t}^{ESS}$
		\ENDIF
		\STATE [$r^{DSM}_{t}$,$r^{ESS}_{t}$] = Reward (s,$a_{t}^{DSM}$,$a_{t}^{ESS}$)
		\STATE Update $Q$, $t$, $SOC$ and $\vec w$   
		\ENDFOR
		\ENDFOR
	\STATE \textbf{Output:}Optimal action sequence from $t=1$ to $T$	
	\end{algorithmic}
\end{algorithm}

\section{Simulation Result}
\subsection{Parameter Settings }

We use MATLAB for our simulations. We model the PV power and the crucial load as shown in Fig.2, which are extracted from \cite{b17}. Surplus PV power is available when PV power is higher than crucial load. Fig.3 shows the Time-of-Use (TOU) electricity price,  which is generated according to winter TOU price of Ontario, Canada. 

\begin{figure}[htbp]
\centering
\includegraphics[width=7.5cm,height=4cm]{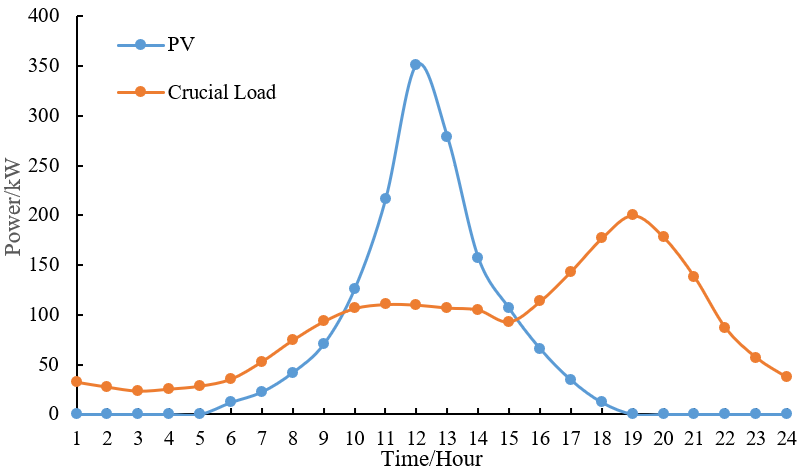}
\caption{PV power and crucial load.}
\label{fig}
\end{figure}

\begin{figure}[!h]
\centering
\includegraphics[width=7.5cm,height=4cm]{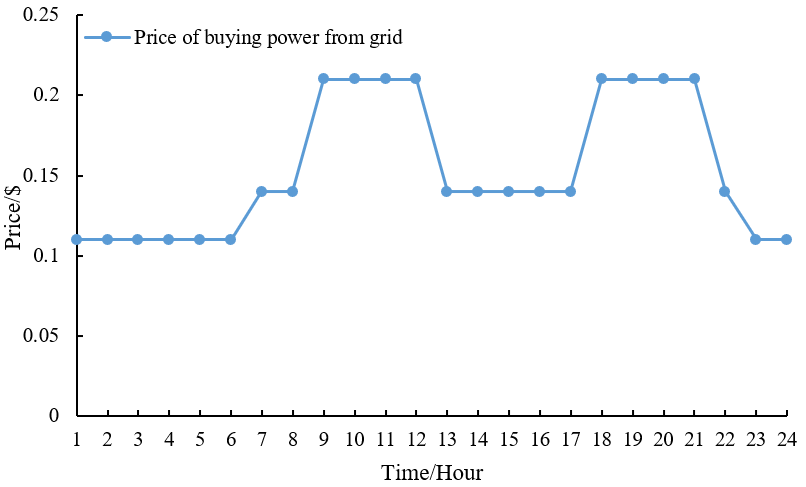}
\caption{TOU electricity price.}
\label{fig}
\end{figure}

\begin{table}[htbp]
\caption{Parameters Settings}
\centering
\renewcommand\arraystretch{1.3}
\begin{tabular}{|p{1.4cm}<{\centering}|p{2.5cm}<{\centering}|p{2.5cm}<{\centering}|}
\hline
\shortstack{Device\\ Number}   &\shortstack{Operation time \\limit (Hours)} & \shortstack{Average duration \\time (Hours) }\\
\hline
1 & [1, 8]& 1\\
\hline
2 & [7, 13]& 1\\
\hline
3& [10, 17]& 2\\
\hline
4& [15, 21]& 1\\
\hline
5& [20, 5$^{\mathrm{+24h}}$]& 3\\
\hline
\end{tabular}
\end{table}

\begin{table}[!h]
\caption{Parameters Settings}
\centering
\renewcommand\arraystretch{1.4}
\begin{tabular}{|p{5.1cm}<{\centering}|p{1.7cm}<{\centering}|}
\hline
Parameters & value\\
\hline
Capacity of ESS: $C^{ESS}$ & 120$kW·h$\\ 
\hline
ESS charge power: $P_{char}$ & 30$kW$\\ 
\hline
Price of PV selling electricity to ESS: $p^{PV}$ & 0.03 $\$/kW·h$ \\
\hline
Price of selling electricity to grid: $p^{sgrid}$& 0.08 $\$/kW·h$ \\
\hline
\end{tabular}
\end{table}

Table 2 shows the operation time limits and the operation duration of five sets of deferrable devices. For each device, the power (kW) is an integer number generated randomly from the set [8, 14] which represents the average power consumed throughout the operation. We repeat the simulations for 30 runs and present the averaged values with 95\% confidence intervals in plots. Table 3 shows the parameter settings of our simulations.

\subsection{Energy Management with Correlated Q-learning}

First, we set the initial SOC of ESS to 0.25, and correlated Q-learning is used to optimize the cost of ESS agent and profit of DSM agent. The result of average DSM cost and ESS profit are shown in Fig.4, where the cost decreases and profit increases through the iterations of the algorithm. It is observed that both agents converge.

\begin{figure}[htbp]
\centering
\includegraphics[width=7.6cm,height=5.7cm]{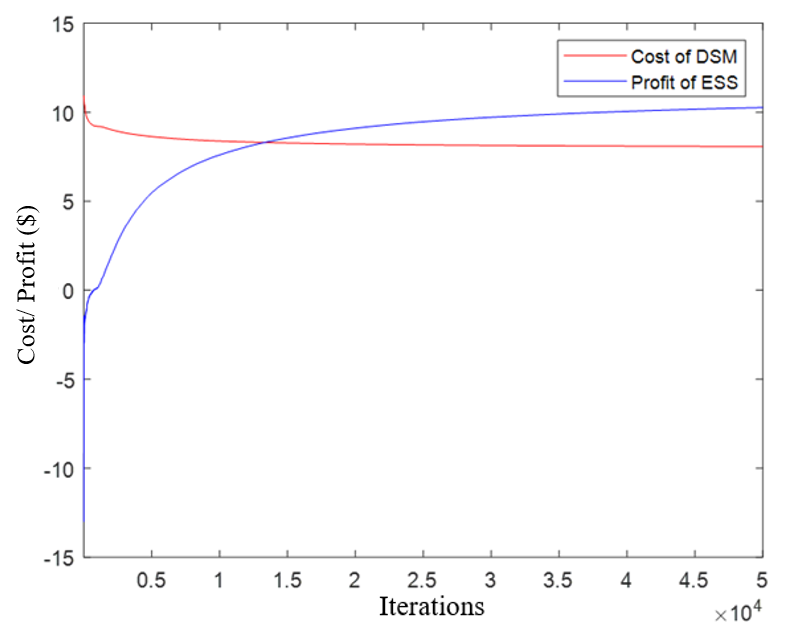}
\caption{Average DSM cost and ESS benefit with iterations.}
\label{fig}
\end{figure}

\begin{figure}[htbp]
\centering
\includegraphics[width=8cm,height=4.7cm]{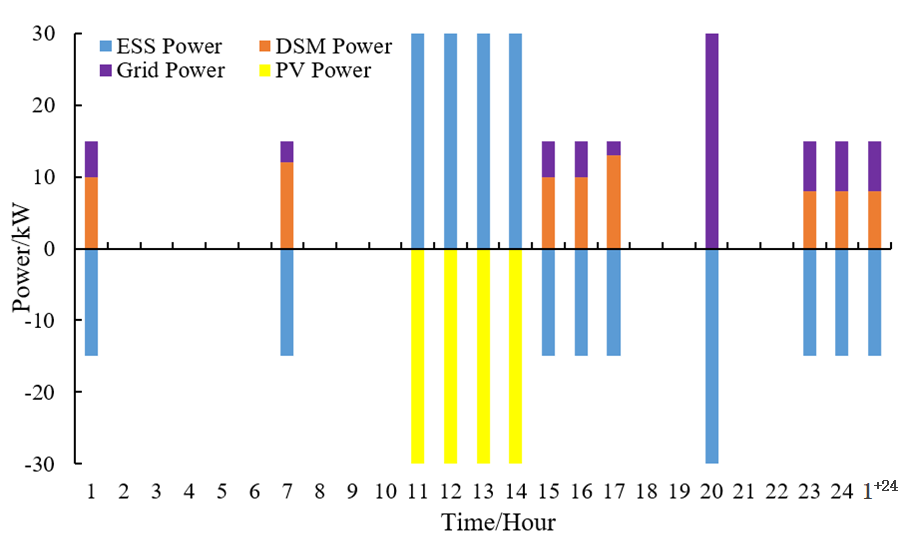}
\caption{Optimal microgrid energy balance under correlated Q-learning}
\label{fig}
\end{figure}

Fig.5 presents how the optimal MG energy management works. Positive bars represent load, charging or electricity sold to grid, while negative bars represent discharging or surplus PV power. The negative bars and positive bars have the same total height at each hour, which means energy balance is maintained. For example, from 1:00 to 2:00, ESS agent discharges with a power 15kW, and DSM agent uses 10kW for operation, while the rest 5kW is sold to main grid. Due to a lower charging price, ESS agent uses surplus PV power to charge from 11:00 to 15:00. Meanwhile, DSM agent uses ESS power for operation, and both agents benefit from cooperation.

 Fig.6 shows how devices are deferred. Device 1 and 2 start immediately  because they have already worked under the minimum TOU price in their time limit. Device 3 and 4 are deferred 5 and 2 hours later respectively for a lower TOU price. Although the TOU price is lower from 13:00 to 18:00, note that device 3 starts operation at 15:00, because the ESS uses surplus PV power for charging from 11:00 to 15:00, and the ESS is unable to discharge in this period. Device 5 are deferred to 1:00 of next day.
 
\begin{figure}[htbp]
\centering
\includegraphics[width=8cm,height=2.5cm]{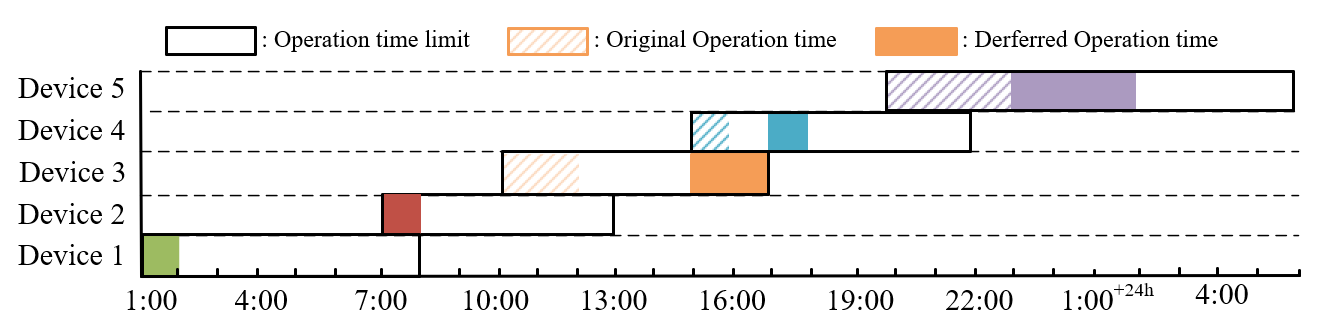}
\caption{Deferred operation time of each device}
\label{fig}
\end{figure}

\subsection{Comparison under Different PV Power}

In the following two sections, we will make a comparison of following three cases: 

Case I: CEQ is used to coordinate two agents. 

Case II: Q-learning without correlation (QLWC) is implemented, where each agent maximizes their own reward and ignore the other agent.

Case III: Q-learning with aggregator (QLWA) is conducted in this case, where all agents belong to one aggregator and maximize the total revenue. Considering the centralized optimization can usually get the global optimal result, we compare CEQ with QLWA to show that the distributed nature of CEQ does not harm total revenue. 

The initial SOC of ESS is set to 0.25, and two cases is compared under different peak PV power. As shown in Fig.7 and Fig.8, the CEQ outperforms QLWC with a lower DSM cost and a higher ESS profit, respectively. The 100\% Peak PV power is defined as Fig.2. At 100\% PV power, CEQ saved 16.5\% cost for DSM agent, and earn 25.5\% more benefit for ESS agent. At 80\% PV power, the proportion for DSM and ESS become 16.3\% and 30.4\%, respectively. CEQ earned at most 44.2\% more benefit than QLWC when peak PV power is only 20\%.  
\begin{figure}[htbp]
\centering
\includegraphics[width=8cm,height=4.5cm]{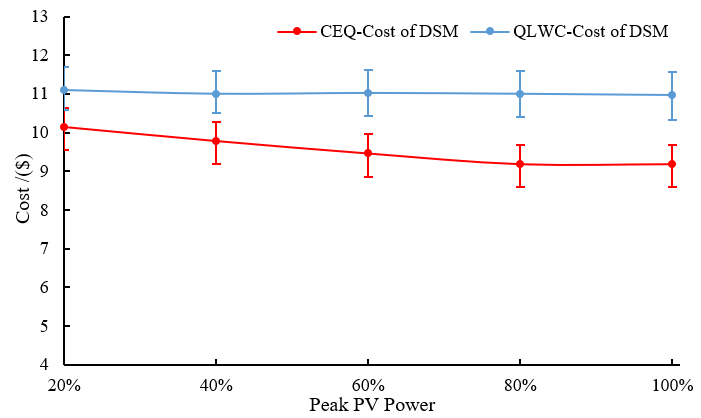}
\caption{Comparison of DSM cost under different PV  power}
\label{fig}
\end{figure}

\begin{figure}[htbp]
\centering
\includegraphics[width=8cm,height=4.5cm]{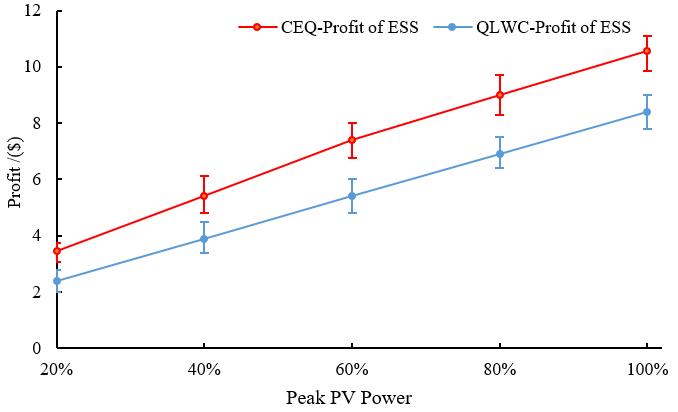}
\caption{Comparison of ESS profit under different PV  power}
\label{fig}
\end{figure}

The main reason is the CEQ is able to coordinate the operation of DSM and ESS. ESS use PV power to charge, and sell the energy to DSM, where both of them benefit from cooperation. On the contrary, in QLWC, both agents try to maximize their own reward and cooperation breaks. Instead of buying energy from ESS agent, DSM agent chooses to buy electricity from main grid. Meanwhile, ESS agent uses PV power to charge, and sell the electricity to main grid.

\subsection{Comparison under Different Initial SOC of ESS}
In this section, we set the PV power to 100\%, and compare three cases under different initial SOC of ESS. Note that the cost of intial SOC is not considered.

\begin{figure}[htbp]
\centering
\includegraphics[width=8cm,height=4.5cm]{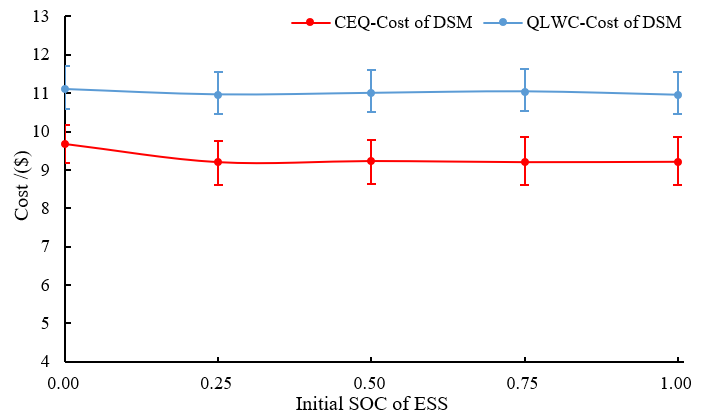}
\caption{Comparison of DSM cost under different initial SOC}
\label{fig}
\end{figure}

\begin{figure}[htbp]
\centering
\includegraphics[width=8cm,height=4.5cm]{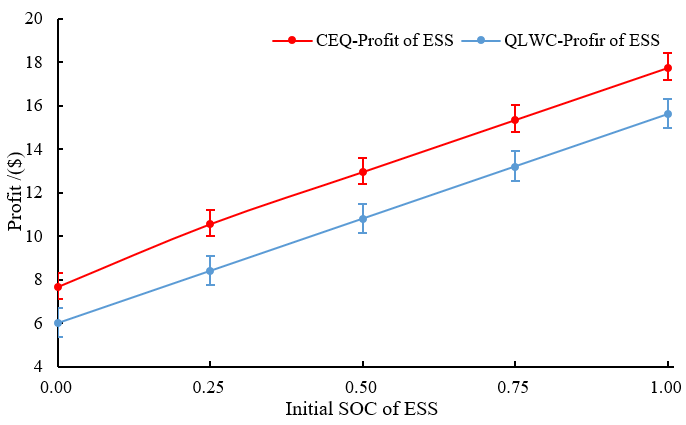}
\caption{Comparison of ESS profit under different initial SOC}
\label{fig}
\end{figure}

As shown in Fig. 9 and Fig. 10, the CEQ has a lower DSM cost and a higher ESS profit than QLWC. DSM agent uses ESS energy for operation, both agents benefit. At 0.75 initial SOC, the DSM agent in CEQ could save 19.3\% cost, and ESS agent in CEQ could earn 16.3\% more benefit than QLWC. After the DSM agent gets enough energy from ESS, the ESS will sell surplus energy to main grid and earn a profit. As a result, the DSM cost remain unchanged, but ESS profit could still increase.

Furthermore, at 100\% PV power, we compare the performance of CEQ with QLWA in Table 4. If we use the profit of CEQ minus the cost, then we will get the same total revenue with QLWA. The result proves CEQ did not harm the total benefit of whole MG. 

\begin{table}[H]
\caption{Comparison of CEQ and QLWC under different initial SOC (\$)}
\centering
\renewcommand\arraystretch{1.4}
\begin{tabular}{|p{1cm}<{\centering}|p{0.9cm}<{\centering}|p{1cm}<{\centering}|p{1cm}<{\centering}|p{1cm}<{\centering}|p{1cm}<{\centering}|}
\hline
\multicolumn{2}{|c|}{Initial SOC} & 0.25& 0.5& 0.75& 1.00\\
\hline
\multirow{2}*{CEQ}&Profit & 10.55& 12.95& 15.35& 17.75\\
\cline{2-6}
&Cost & 9.20& 9.23& 9.20& 9.21\\
\hline
\multicolumn{2}{|c|}{QLWA-Revenue} & 1.35 & 3.72& 6.15 & 8.54\\
\hline
\end{tabular}
\end{table}

\section{Conclusion}
In this paper, we propose a  multi-agent correlated Q-learning algorithm for microgrid energy management. Compared with the Q-learning without correlation, we find that the proposed CEQ scheme could save at most 19.3\% cost for DSM agent, and earn at most 44.2\% more benefit for ESS agent. The simulation results show that CEQ is capable of successfully coordinating the operation of agents. In addition, compared with an aggregator case, CEQ is shown to maintain the same total benefit for the agents.

\section*{Acknowledgment}

This work is supported by NSERC CREATE (497981), NSERC Canada Research Chairs Program and the U.S. National Science Foundation (NSF) under Grant Number CNS-1647135.

\end{document}